\documentclass[a4paper]{article}

\usepackage{color}
\usepackage{hyperref}

\usepackage[bitstream-charter]{mathdesign}
\DeclareSymbolFont{usualmathcal}{OMS}{cmsy}{m}{n}
\DeclareSymbolFontAlphabet{\mathcal}{usualmathcal}

\begin{document}

\begin{center}{\Large \textbf{Classical and quantum gravity from relativistic quantum mechanics}}\end{center}

\begin{center}
Walter Smilga
\end{center}

\begin{center}
Geretsried, Germany
\\
wsmilga@compuserve.com
\\
\end{center}

\begin{center}
\today
\end{center}

\iffalse
\definecolor{palegray}{gray}{0.95}
\begin{center}
\colorbox{palegray}{
  \begin{tabular}{rr}
  \begin{minipage}{0.1\textwidth}
    \includegraphics[width=20mm]{logo_ICGTMP.png}
  \end{minipage}
  &
  \begin{minipage}{0.85\textwidth}
    \begin{center}
    {\it XII. International Symposium on Quantum Theory and Symmetries (QTS12), Prague, 24 - 28 July 2023}\\
 
    \end{center}
  \end{minipage}
\end{tabular}
}
\end{center}
\fi

\section*{Abstract}
{\bf

It is common practice to describe elementary particles by irreducible unitary representations of the Poincar\'e group.
In the same way, multi-particle systems can be described by irreducible unitary representations of the Poincar\'e group.
Representations of the Poincar\'e group are characterised by fixed eigenvalues of two Casimir operators corresponding to a fixed mass and a fixed angular momentum.
In multi-particle systems (of massive spinless particles), fixing these eigenvalues leads to correlations between the particles.
In the quasi-classical approximation of large quantum numbers, these correlations take on the structure of a gravitational interaction described by the field equations of conformal gravity.
A theoretical value of the corresponding gravitational constant is calculated. 
It agrees with the empirical value used in the field equations of general relativity.}
 
%\vspace{10pt}
%\noindent\rule{\textwidth}{1pt}
%\tableofcontents\thispagestyle{fancy}
%\noindent\rule{\textwidth}{1pt}
%\vspace{10pt}

%\newpage
 
\section{Introduction}
\label{sec:intro}

Although the axioms of (non-relativistic) quantum mechanics have long since found a generally accepted form, their validity for relativistically interacting multi-particle systems has not yet been seriously considered.
The general view is still that such systems must be described by local quantum field theories\footnote{Werner Heisenberg in his last lecture on elementary particle physics (1965): `Today we know that a theory of elementary particles must be a local quantum field theory.'}.
This is why we now rely on a model based on interacting quantised fields, the Standard Model of particle physics, which describes three of the four fundamental interactions.
Before the development of non-perturbative methods (e.g.\,lattice QCD), numerical results from this model were obtained using perturbation algorithms. 
Unfortunately, these algorithms lead to mathematical inconsistencies at higher orders.
These inconsistencies have prevented a full understanding of the mathematical and physical content of the model. 
Instead, they have opened the door to fanciful, reifying interpretations \cite{dm} (`virtual particles constantly popping in and out of existence'\footnote{Scientific American, 9 October 2006}), which have further blocked the way to a rational understanding of the model.
Whether the purely numerical lattice methods will lead to a deeper understanding remains to be seen. 

If by understanding we mean a mathematically consistent traceability to the two fundamental principles of nature, as formulated by the axioms of quantum mechanics and Poincar\'e invariance, then we certainly have not yet understood the Standard Model.
After decades of stagnation on this issue, it seems that only an unbiased, model-independent view of relativistic particle physics can lead us out of this impasse.

A promising perspective emerges from the following observation:
Just like linking two angular momenta to a resulting angular momentum, linking two single-particle states, each belonging to an irreducible representation of the Poincar\'e group, to a physically meaningful two-particle state is by no means a trivial task. 
It is not sufficient to simply form the product state of two single-particle states; the representation theory of the Poincar\'e group rather tells us, that the product representation obtained this way must be reduced to a direct sum of irreducible representations in order to obtain descriptions of two-particle systems with well-defined linear and angular momenta.
The resulting two-particle states are then obtained as superpositions of product states with coefficients equal to the Clebsch--Gordan coefficients of the Poincar\'e group \cite{hj}.
The Standard Model, although based on product representations, clearly ignores this rule.

This article examines the consequences of strict adherence to this rule.

\section{Axioms of relativistic quantum mechanics}

Dirac's axioms of quantum mechanics \cite{pamd} can be formulated as follows\footnote{from lecture notes http://web.mit.edu/8.05/handouts/jaffe1.pdf}:

\noindent
{\bf First Postulate}: {\it At each instant the state of a physical system is represented by a normalised ket
$\left.\mbox{\textbar}\psi \right>$ in the space of states.}

\noindent
{\bf Second Postulate}: {\it Every observable attribute of a physical system is described by an operator that acts on the kets that describe the system.}

\noindent
{\bf Third Postulate}: {\it The only possible result of the measurement of an observable $\mathcal{A}$ is one of the eigenvalues of the corresponding operator $\hat{\mathcal{A}}$.}

\noindent
{\bf Fourth Postulate}: {\it When a measurement of an observable $\mathcal{A}$ is made on a given normalised state $\left.\mbox{\textbar}\psi \right>$, the probability of obtaining an eigenvalue $a_n$ is given by the square of the inner product of $\left.\mbox{\textbar}\psi \right>$ with the normalised eigenstate 
$\left.\mbox{\textbar}a_n\right>$, $|{\left<a_n|\psi \right>}|^2$.}

\noindent
{\bf Fifth Postulate}: {\it Immediately after the measurement of an observable $\mathcal{A}$ has yielded a value $a_n$, the state of the system is the normalised eigenstate $\left.\mbox{\textbar}a_n\right>$.}

\noindent
{\bf Sixth Postulate}: {\it 
The time evolution of the state of a quantum system is described by\\ 
$\left.\mbox{\textbar}\psi(t)\right> = \hat{U}(t,t_0)\left.\mbox{\textbar}\psi(t_0)\right>$, for some unitary operator $\hat{U}$.} 

Note that the sixth axiom is only valid as long as no measurement is made according to the fifth axiom.
Ignoring this restriction leads leads to the so-called `measurement problem'.

In this general form, these axioms do not distinguish between non-relativistic and relativistic particle systems.

If $\left.\mbox{\textbar}\psi\right>$ belongs to a product representation of the Poincar\'e group, it is a trivial exercise to write down the corresponding operator $\hat{U}(t,t_0)$ (see, e.g.,\cite{sss}, pp. 148--150).
Of course, a product representation only describes `free', i.e. non-interacting particles.

Historically, the challenge has been to add the four fundamental interactions in a way that preserves relativistic covariance.
This led to the Standard Model of particle physics. 
Instead of relying on the results of the representation theory of the Poincar\'e group \cite{ew}, this model explicitly models interactions by additional interaction-mediating fields. 
It redefines $\left.\mbox{\textbar}\psi\right>$ as an operator-valued field based on a Fock-space representation \cite{vaf}, misleadingly called `second quantisation'. 
When applied to a `vacuum' state $\left.\mbox{\textbar}0\right>$, this field operator generates a multi-particle state.
The operator $\hat{U}(t,t_0)$ is then derived from a Lagrangian based on the matter field (fermion field) and a bosonic field that provides the interaction, both linked by an interaction term.  
Transition amplitudes are determined using a perturbation algorithm in which the interaction term is treated as a (small) perturbation of the matter field.

Unfortunately, this perturbation algorithm leads to mathematical inconsistencies that can only be solved by a formal trick called `renormalisation'.
Since this trick does not work for the gravitational interaction, the Standard Model describes only the electromagnetic, weak, and strong interactions.

This article takes an alternative, non-perturbative approach.
It is firmly based on the representation theory of the Poincar\'e group.
After Wigner's seminal paper in 1939 \cite{ew}, elementary particles have been successfully described by irreducible unitary representations of the Poincar\'e group.
It is therefore only logical to also describe multi-particle configurations of elementary particles by irreducible unitary representations of the Poincar\'e group and to consider $\left.\mbox{\textbar}\psi\right>$ as a state of such a representation.
Under this premise, the above axioms can be regarded as axioms of relativistic quantum mechanics.

\section{Irreducible representations of the Poincar\'e group}
\label{sec:1}

Irreducible unitary representations of the Poincar\'e group $P(3,1)$ are characterised by fixed eigenvalues of two Casimir operators, corresponding to a mass and an intrinsic angular momentum. 
This is analogous to the mass and spin of irreducible single-particle representations (see textbooks on quantum field theory, e.g.\cite{sss}, pp. 36--53). 
The state space of such a representations has an orthonormal basis consisting of eigenstates of a complete set of commuting operators, which, in addition to the two Casimir operators, can be the four components of the (total) momentum and a component of the pseudovector $\frac{1}{2}\epsilon_{\sigma\mu\nu\lambda} M^{\mu\nu}p^\lambda$, which in the rest frame corresponds to a component of the angular momentum $M^{ik}$. % (see \cite{sss}, p.\,45).
In contrast to single-particle representations, where the spin determines the angular momentum, in multi-particle representations an intrinsic orbital angular momentum also contributes to the total angular momentum.

The reduction of a product representation to a sum of irreducible representations implies a decomposition of the associated state space, which in turn leads to correlations between the individual eigenstates of momentum $\left.\mbox{\textbar}\mathbf{p}_1\right>$ and $\left.\mbox{\textbar}\mathbf{p}_2\right>$:
Fixing the eigenvalue of the first Casimir operator correlates the particle states in such a way that the mass, defined by the total momentum $\mathbf{p} = \mathbf{p}_1 + \mathbf{p}_2$, takes on a fixed value.
Fixing the eigenvalue of the second Casimir operator forces the particle states into a rotationally symmetric momentum-entangled structure, such as in the two-particle eigenstate of the total linear and angular momentum (see \cite{ws})
\begin{equation}
\left.\mbox{\textbar}\mathbf{p},w \right> = \int_\Omega \! \mathrm{d}\omega_\mathbf{k}
\left.\mbox{\textbar}\mathbf{p}_1+\mathbf{k} \right> 
c_{\mathbf{p},w}(\mathbf{k})  
\left.\mbox{\textbar}\mathbf{p}_2-\mathbf{k} \right>.    		 \label{5-15}
\end{equation}  
The integration over $\mathbf{k}$ is over a circular path $\Omega$.
This structure can be understood as description of two particles correlated (entangled) by the virtual exchange of momenta $\mathbf{k}$.

Integrals with a similar structure can be found in the perturbation algorithm of the Standard Model.
There, however, the integration is performed over the entire three-dimensional momentum space.
For loop-like Feynman diagrams, this leads directly to the well-known divergences.

The two-particle state (\ref{5-15}) clearly shows that the particle momenta cannot (not even approximately) be eigenstates simultaneously with the total linear momentum and angular momentum.
This property already follows from the commutation relations of the Poincar\'e group:
In fact, the commutation relations of the infinitesimal generators of $P(3,1)$ (see \cite{sss}, p.\,45) allow the simultaneous diagonalisation of a linear momentum (the total momentum) and of an angular momentum component with an axis of rotation parallel to the linear momentum; they do not, however, allow the simultaneous diagonalisation of another momentum that is not parallel to the axis of rotation.

The rotational symmetry of the eigenstates of the orbital angular momentum is preserved in the quasi-classical limit of large quantum numbers.
The corresponding wave functions define probability distributions in the form of well-defined rotationally symmetric closed orbits (see \cite{ed}, pp.\,27--29). 
Wave packets travelling along such orbits necessarily have time-dependent momenta that are subject to the conservation of the total momentum.

\section{Conservation of total momentum in multi-particle systems}
\label{sec:2}

In the quasi-classical limit, the conserved total momentum of the multi-particle system under consideration can therefore be written as a sum over time-dependent particle momenta.
In the rest frame of the system, the conservation law of momentum takes the form
\begin{equation}
\sum_{i=1}^N \mathbf{p}_i (t) = \mathbf{0} .							\label{2-3}
\end{equation}
The parameter $t$ denotes the time of a time cut.

In Newtonian mechanics, a time-dependent particle momentum $\mathbf{p}_i (t)$ defines a force that is equal to the time derivative of that momentum, whatever the reason for the time dependence.
Since the total momentum is conserved, this force must be compensated by forces from other particles.
Thus we can describe the time dependence of the particle momenta in terms of an exchange of forces or, equivalently, by a virtual (instantaneous) exchange of momentum between the particles.

Note that `virtual exchange of momentum' is used here as a convenient descriptive phrase for describing the structure of Eq.\,(\ref{2-3}) and its consequences. 
It is not meant in the sense of a real exchange of momentum, which would be subject to relativistic causality.

This description is valid for arbitrary particle momenta and masses.
It is therefore invariant under rotations and boost operations applied to the individual particle momenta.
Furthermore, it is invariant under conformal scaling of the individual particle momenta, i.e.\,individual scaling of the particle masses.
These properties define the intrinsic symmetries of Eq.\,(\ref{2-3}).

If Eq.\,(\ref{2-3}) is rewritten in the form 
\begin{equation}
\mathbf{p}_1 (t) = - \sum_{i=2}^N \mathbf{p}_i (t),			\label{2-4}
\end{equation}
it is then clear that the trajectory of particle~$1$ is completely determined by the time-dependent momenta of particles $2$ to $N$, together with the position and momentum of particle~$1$ at $t = 0$. 
Had we not known that we were dealing with the conservation of total momentum, we might have suspected that the mere presence of neighbouring matter would bend the trajectory of particle~$1$.
Alternatively, we could have argued that the presence of matter bends spacetime along the trajectory of  particle~$1$.
This clearly evokes associations with the phenomenon of gravitation and suggests that particles $2$ to $N$ generate a virtual gravitational field, expressed by the metric of spacetime, which bends the trajectory of particle~$1$ in such a way that the total momentum of the system is the same for all times $t$.

Despite the virtual nature of this field, its structure is in principle measurable by observing the trajectories of particle~$1$.
A structure that can be measured should also be theoretically describable.
Of course, this description must be compatible with Einstein's principle of relativity.

\section{Relativistic field equations}
\label{sec:4}

In establishing the field equations of general relativity, Einstein required that `the general laws of nature must be expressed by equations that are valid for all coordinate systems, that is, that are covariant (generally covariant) with respect to arbitrary substitutions' (\cite{ae}, p.\,776, translation by the author). 
We adopt this requirement and also the coordinate-independent description of curved trajectories in four-dimensional spacetime with the tools of differential geometry. 

The concept of virtual momentum exchange is easily brought into a four-dimen\-sional form by replacing $\mathbf{p}_i (t)$ by a momentum density $\mathbf{p}_i (x)$ and combining $\mathbf{p}_i (x)$ and the corresponding energy density $p^0_i(x)$ into the four-vector $p_i (x)$. 
As required by Einstein, $p_i (x)$ then becomes part of a covariant energy-momentum tensor $T^{\mu\nu}(x)$ (of massive point-like particles, see, e.g.,\cite{apgn}).

The variations of the momentum along a trajectory `generate' virtual exchange momenta.
Since these exchange momenta are derived from the time-dependent particle momenta alone, they are independent of the trace $T$ of $T^{\mu\nu}$ corresponding to the (time-constant) particle masses.
At the same time, other particles `absorb' the exchange momenta, causing their trajectories to change. 
These changes can again be described in terms of time-dependent momenta or, preferably, in terms of the curvatures of the trajectories, which are easier observed by a distant observer. 
Again, only the traceless part of $T^{\mu\nu}$ is involved, namely
\begin{equation}
T^{\mu\nu} - \textstyle{\frac{1}{4}} g^{\mu\nu}\,T.   \label{4-5}
\end{equation}
This is consitent with the fact that the masses of the particles do not enter into the conservation law of momentum, Eq.\,(\ref{2-3}). 

The exchange process will now be formulated as a law of nature (in the sense of Einstein's requirement); that is, as a covariant relation between the traceless energy-momentum tensor (emitter) and a geometric tensor $W^{\mu\nu}$ (absorber), derived from a curvature tensor.
Together, these tensors are expected to form the covariant equation
\begin{equation}
W^{\mu\nu} = \kappa\,(T^{\mu\nu} - \textstyle{\frac{1}{4}} g^{\mu\nu}\,T) ,			\label{4-7}
\end{equation}
where $\kappa$ is a constant needed for dimensional reasons. 
As a counterpart to the traceless energy-momentum tensor, $W^{\mu\nu}$ must also be traceless.

To determine the explicit form of $W^{\mu\nu}$, we follow Mannheim \cite{pm} and use the variational principle of least action $\delta I = 0$.
This requires a suitably chosen action $I = I_W + I_M$, the sum of a geometric action $I_W$ and a material action $I_M$, which is invariant under all relevant coordinate transformations, including local conformal scaling.
The functional variation of $I$ with respect to the metric tensor $g^{\mu\nu}$ should then give the expected covariant relation between the curvature tensor and the energy-momentum tensor. 

The only geometric action that is invariant under local conformal scaling is the Weyl action 
(see \cite{pm}) 
\begin{equation}
I_W =  - \textstyle{\frac{1}{4}} \int\!\mathrm{d}^4x\,(-g)^{1/2}\,C_{\lambda \mu \nu \xi} C^{\lambda \mu \nu \xi}, \label{4-8}
\end{equation}
where $C_{\lambda \mu \nu \xi}$ is the conformally invariant Weyl tensor, which is the traceless part of the Riemann curvature tensor $R_{\lambda\mu\nu\xi}$.

For the material action,
\begin{equation}
I_M = \kappa \int\!\!\mathrm{d}^4x\,(-g)^{1/2}\,g_{\mu\nu}\,(T^{\mu\nu} - \textstyle{\frac{1}{4}} g^{\mu\nu}\,T) \label{4-9}
\end{equation}
can be chosen.
The term under the integral is the trace of the traceless energy-momentum tensor, which is zero in any coordinate system:
$g_{\mu\nu}\,(T^{\mu\nu} - \textstyle{\frac{1}{4}} g^{\mu\nu}\,T)\ =\\g_{\mu\nu}\,T^{\mu\nu} - T = T - T = 0$.
Thus $I_M$ is trivially invariant under local conformal scaling.

The factor $(-g)^{1/2}$ in the integrals $I_W$ and $I_M$ can be dropped if $x$ refers to a fixed Minkowskian coordinate system (see \cite{ae}, p.\,801).

As shown in \cite{pm}, the functional derivative of $I_W + I_M$ with respect to $g^{\mu\nu}$ yields the fourth-order partial differential equations 
\begin{equation}
		W^{\mu\nu} = \kappa\, (T^{\mu\nu} - \textstyle{\frac{1}{4}} g^{\mu\nu}\,T),		\label{4-13}
\end{equation} 
where $W^{\mu\nu}$, expressed in terms of the Riemann tensor and Ricci tensor $R^{\mu\nu}$, has the form
\begin{eqnarray}
W^{\mu\nu} &=& \frac{1}{2} g^{\mu\nu}(R^\alpha_{\;\;\alpha})^{;\beta}_{\;\;\;;\beta} + R^{\mu\nu;\beta}_{\;\;\;\;\;\;\;;\beta}
- R^{\mu\beta;\nu}_{\;\;\;\;\;\;\;;\beta} - R^{\nu\beta;\mu}_{\;\;\;\;\;\;\;;\beta} - 2 R^{\mu\beta} R^\nu_{\;\;\beta} 
\hspace{1.7cm} \nonumber \\  
&+& \frac{1}{2} g^{\mu\nu} R_{\alpha\beta} R^{\alpha\beta} - \frac{2}{3} g^{\mu\nu} (R^\alpha_{\;\;\alpha})^{;\beta}_{\;\;\;;\beta}
+ \frac{2}{3}(R^\alpha_{\;\;\alpha})^{;\mu;\nu} + \frac{2}{3}R^\alpha_{\;\;\alpha} R^{\mu\nu} \nonumber  \\
&-& \frac{1}{6} g^{\mu\nu} (R^\alpha_{\;\;\alpha})^2    \nonumber \\
&=&  2 \,C^{\mu\lambda\nu\xi}_{\;\;\;\;\;\;\;\;\;\;\;;\lambda;\xi} - C^{\mu\lambda\nu\xi} R_{\lambda\xi}. 					\label{4-14}
\end{eqnarray}

Due to their derivation by a variational principle, these field equations describe an equilibrium between emitted and absorbed virtual momenta.
Surprisingly, these equations have the same structure as Mannheim's field equations of conformal gravity, although the latter are based on completely different physical premises.
While Mannheim postulated a gauge invariance of nature under local conformal scaling, conformal covariance here refers to the aforementioned property of the conservation law of momentum to be covariant under particle-individual conformal scaling.
In contrast to Mannheim's approach, the particles involved here need not be massless.

As in Einstein's general theory of relativity, the metric tensor $g^{\mu\nu}(x)$ defines a pseudo-Riemannian manifold. 
Observers moving freely along a trajectory in this manifold will experience a curved spacetime, whereas they will infer the existence of a gravitational field from a Minkowskian coordinate system.

Mathematically, the field equations are inhomogeneous differential equations with the traceless energy-momentum tensor as the source (of virtual momenta). 
They exploit the equivalence between the description of trajectories in terms of time-dependent momenta -- that is, by $T^{\mu\nu} - \textstyle{\frac{1}{4}} g^{\mu\nu}\,T$ -- and their description in terms of curvatures -- that is, by $W^{\mu\nu}$.

Since these are differential equations, a virtual momentum generated by a particle at position $x$ can be absorbed by a second particle at or near position $x$, but also by particles at other positions.
Therefore, only a fraction of the virtual momentum will be absorbed by the second particle.
Since Eq.\,(\ref{2-3}) does not favour any particle, this must also apply to the field equations. 
Quantum mechanically, this means that all particles have the same initial probability of being involved in the exchange process.
Therefore, $\kappa$ must contain a constant probability factor of $1/N$, where $N$ is the total number of particles.
This gives $\kappa$ the role of a coupling constant (see next section).

When applied to a real multi-particle system (subject to the conservation law of momentum), the field equations translate the dynamical constraints on the particle momenta, i.e.\,the conservation of total momentum expressed by Eq.\,(\ref{2-3}), into equivalent kinematic constraints on the particle trajectories in spacetime.
Thus, the field equations describe the structure of spacetime of isolated multi-particle systems when the energy-momentum tensor is given, and vice versa.
As Wheeler put it in \cite{jw} (slightly modified by the author): Matter tells spacetime how to curve; spacetime tells matter how to move.

\section{Coupling constants}
\label{sec:5}

The constant $\kappa$ in the field equations (\ref{4-13}) corresponds to the gravitational constant in Einstein's field equations.
It connects the energy-momentum tensor, which has the dimension of an energy density, i.e. $\mathbf{J\,m^{-3}}$ or equivalently $\mathbf{kg\,m^{-1} s^{-2}}$, with a squared curvature, which has the dimension $\mathbf{m^{-4}}$.
Therefore, $\kappa$ has the dimension $\mathbf{kg^{-1} m^{-3} s^{2}}$ or equivalently $\mathbf{J^{-1}\,m^{-1}}$.
It converts the dimension of the energy into the dimension of an inverse length, which can be understood as the dimension of the curvature of a trajectory.

Remarkably, there are two constants of nature, which together have the same dimension as $\kappa$:
The Planck constant $h$ with the dimension $\mathbf{J\,s}$ and the speed of light $c$ with the dimension $\mathbf{m\,s^{-1}}$  form the constant $1/(h\,c)$, which has the dimension $\mathbf{J^{-1}\,m^{-1}}$.
Its numerical value is $1 / (6.626 \times 10^{-34} \times 2.998 \times 10^{8})$ $= 5.034 \times 10^{24}$.

Together with the dimensionless factor $1/N$, where $N$ is tentatively $2.4 \times 10^{67}$, the estimated number of atoms in our galaxy\footnote{Steve Cavil on Oxford Education Board:
``Our galaxy, the Milky Way, contains approximately 100 to 400 billion stars. If we take this as 200 billion or $2 \times 10^{11}$ stars and assume that our sun is a reasonable average size we can calculate that our galaxy contains about $(1.2 \times 10^{56}) \times (2 \times 10^{11}) = 2.4 \times 10^{67}$ atoms.''},
we get an estimate of $\kappa$: 
\begin{equation}
\kappa = 1 / (N\,h\,c) \approx 2.1 \times10^{-43}\;\mathbf{kg^{-1} m^{-3} s^{2}}.  \label{5-1}
\end{equation}

For our solar system the equations of conformal gravity are equivalent to those of general relativity (see \cite{pm}).
This estimate can therefore be compared to the value of the gravitational constant in Einstein's field equations.
Based on the measured value of Newton's gravitational constant $G = 6.674 \times 10^{-11}\;\mathbf{kg^{-1} m^3 s^{-2}}$, this value is 
\begin{equation}
\kappa_E = 8\pi G/c^4 \approx 2.077 \times 10^{-43}\;\; \mathbf{kg^{-1} m^{-1} s^2}.                 \label{5-2}
\end{equation}

The fact that $\kappa$ can be expressed in terms of $h$, $c$, and the dimensionless constant $N$ with a unique origin removes the motivation to speculate about physics at Planck scales.
The factor $1/N$ makes clear that the gravitational interaction, unlike the electromagnetic interaction, is a multi-particle interaction, which has an extremely large range due to the lack of shielding.
This factor also gives a simple answer to the so-called hierarchy problem.

For the electromagnetic interaction, the physically relevant representation is an irreducible two-particle representation, since in larger systems positively and negatively charged particles tend to combine into a configuration that is electrically neutral to the outside.

In the article \cite{ws} cited above, the structure of irreducible (massive) two-particle representations was analysed leading to the correct value for the electromagnetic fine-structure constant $\alpha$
 (= 1/137.03608245 vs.\,CODATA value $1/137.035999139$).
The agreement of the calculated value with the experimental value identifies the kinematic constraints of irreducible two-particle representations as `electromagnetic interaction' and thus provides further support for the concept of irreducible multi-particle representations of the Poincar\'e group.

\section{Quantisation of spacetime}
\label{sec:7}

As described above, irreducible multi-particle representations of the Poincar\'e group in the limit of large quantum numbers lead to curved trajectories in a pseudo-Riemannian spacetime.
The inversion of this limit can therefore be understood as `quantisation of spacetime'.
This means that continuous spacetime is replaced by discrete trajectories, which are then replaced by wave functions of an irreducible multi-particle representation of $P(3,1)$.
This quantisation rule is transparent and well founded, both physically and mathematically. 
Unlike other attempts to quantise spacetime, such as loop quantum gravity or string theory, it provides a quantisation of spacetime at experimentally accessible scales.

\section{Conclusions}
\label{sec:8}

This article provides a consistent foundation for the phenomenon of gravity in both the quantum mechanical and classical domains.
The foundation is based solely on the first principles of quantum mechanics and Poincar\'e invariance, unified in the representation theory of the Poincar\'e group.

In multi-particle systems, which are described quantum mechanically by irreducible multi-particle representations of the Poincar\'e group, the kinematics of the individual particles is constrained in the sense that the individual particle states are superposed in such a way that they form a common rotationally symmetric eigenstate of the orbital angular momentum. 
As a consequence, in the quasi-classical limit, the trajectories of the particles are not rectilinear, but form a pseudo-Riemannian manifold, which is described by the field equations of conformal gravity (CG).

The field equations of CG are currently being discussed as an alternative to the field equations of general relativity (GR) (see e.g.\,\cite{pm}).
In the form presented here, they apply to isolated multi-particle systems consisting of massive spinless particles or macroscopic bodies. 
For the solar system, they not only reproduce the Schwarzschild solutions of GR  but (in contrast to GR) they are also able to correctly describe galactic rotation curves without the aid of exotic dark matter (see \cite{pm} and the references given there).
Thus, not only Einstein--Newton gravity but also the puzzling galactic rotation curves can be regarded as manifestations of the constrained kinematics of isolated multi-particle systems.

Applied to our galaxy, the derived field equations provide a gravitational constant of the right order of magnitude. 
The numerical agreement of this constant with Einstein's gravitational constant confirms that the strength of gravity in our solar system is indeed determined by the surrounding Milky Way galaxy.
This can be generalised to the statement that the gravitational constant within a galaxy is specific to that galaxy. 

If we agree to define quantum gravity as the description of multi-particle systems that is consistent with the axioms of quantum mechanics and leads in the quasi-classical limit to the field equations of CG, then irreducible unitary multi-particle representations of the Poincar\'e group are the answer to the current search for a quantum theory of gravity.
These representations are in principle well-understood and provide the simplest description that fits this definition.

In summary, the above considerations show that it is possible to describe relativistically interacting multi-particle systems on the basis of the axioms of quantum mechanics, without having to resort to the problematic methods of interacting quantum fields.
The absence of explicit interaction terms means that the mathematical complexity is drastically reduced compared to perturbation schemes based on quantum field theories as known from the Standard Model.
A reformulation of the Standard Model based on irreducible multi-particle representations of the Poincar\'e group would not only allow a smooth integration of gravity into the model, but would also eliminate renormalisation problems.
The main advance over the current Standard Model is that this approach not only avoids the inconsistencies of local field theories, but also provides exactly the two known long-range interactions, together with their coupling constants.

Of course, these results do not devalue the Standard Model as a collection of powerful algorithms ready for practical computation.
But they do call into question the role of gauge invariance as the cause of interactions.
And they expose the myth of `virtual particles constantly popping in and out of existence' as merely an inappropriately reifying interpretation of the Fock space-based perturbation algorithm or, in the words of A.~N.~Whitehead, as a `fallacy of misplaced concreteness' \cite{anw}.

%\section*{Acknowledgements}

\paragraph{Funding information}
%\label{sec:9}

No funds, grants, or other support were received.

%\nolinenumbers

\end{document}